\documentclass[12pt]{article}
\usepackage{epsfig,amssymb}
\usepackage{latexsym}

\newcommand{\bq}{\begin{equation}}
\newcommand{\eq}{\end{equation}}
\newcommand{\bqa}{\begin{eqnarray}}
\newcommand{\eqa}{\end{eqnarray}}
\newcommand{\ben}{\begin{enumerate}}
\newcommand{\een}{\end{enumerate}}
\newcommand{\bc}{\begin{center}}
\newcommand{\ec}{\end{center}}
\newcommand{\bqb}{\begin{eqnarray*}}
\newcommand{\eqb}{\end{eqnarray*}}

\def\pr#1#2#3{ Phys. Rev. ${\bf{#1}}$,#2 (#3)}

\def\pl#1#2#3{ Phys. Lett. ${\bf{#1}}$,#2 (#3)}
\def\prep#1#2#3{ Phys. Rep. ${\bf{#1}}$,#2 (#3)}

\def\np#1#2#3{ Nucl. Phys. ${\bf{#1}}$,#2 (#3)}

\def\ie{{\it i.e. ~}}
\def\eg{{\it e.g. ~}}

\def\etal{{\it et.al.~}}

\global\nulldelimiterspace = 0pt

\def\mwd{m_W^2}

\def\mzd{m_Z^2}

\begin{document}
\pagenumbering{arabic}
\thispagestyle{empty}
\def\thefootnote{\fnsymbol{footnote}}
\setcounter{footnote}{1}


\begin{center}
Short version of the paper hep-ph/0207273\\
November 2002\\
to appear in Phys.Rev.D\\

 \end{center}
\vspace{2cm}
\begin{center}
{\Large\bf The processes
$e^-e^+\to\gamma\gamma, Z \gamma, ZZ$ in SM and
MSSM\footnote{Programme d'Actions Int\'egr\'ees Franco-Hellenique,
 Platon 04100 UM}}
 \vspace{1.5cm}  \\
{\large G.J. Gounaris$^a$, J. Layssac$^b$  and F.M. Renard$^b$}\\
\vspace{0.2cm}
$^a$Department of Theoretical Physics, Aristotle
University of Thessaloniki,\\
Gr-54124, Thessaloniki, Greece.\\
\vspace{0.2cm}
$^b$Physique
Math\'{e}matique et Th\'{e}orique,
UMR 5825\\
Universit\'{e} Montpellier II,
 F-34095 Montpellier Cedex 5.\\

\vspace*{1.cm}

{\bf Abstract}
\end{center}
We present  the results of a complete analysis of the one loop
electroweak corrections to  $e^-e^+\to\gamma\gamma, ~Z\gamma, ~ZZ$
in the Standard (SM)  and the Minimal Supersymmetric Standard
Model (MSSM). A special emphasis is put on the high energy behaviour
of the various helicity amplitudes and the way the logarithmic
structure is generated. The large magnitude of these effects,
which induce striking differences between the SM and MSSM cases
at high energies, offers the possibility of making global tests
which could check the consistency of these models,
and even decide  whether any additional new physics
is required.

\vspace{0.5cm}
PACS 12.15.-y, 12.15.Lk, 14.70.-e

\def\thefootnote{\arabic{footnote}}
\setcounter{footnote}{0}
\clearpage

\section{Introduction}

Several projects of high energy and high luminosity $e^-e^+$
colliders (LC, CLIC) are under consideration \cite{LC, CLIC},
possibly with a photon-photon option \cite{ggcoll}.
They should allow not only to produce new particles but also,
to make very precise tests of the fundamental interactions.\par
It is by now well-known that the electroweak radiative corrections
to several standard processes strongly increase with the energy.
This arises due to the presence already at the   1-loop level,
of large double (DL) and single (SL) logarithm terms behaving like
$(\alpha/\pi) \ln^2s$,  $(\alpha/\pi) \ln s$,
\cite{Sud1, Sud2, log}. In the TeV range such terms reach the
several tens of percent level.
They are no more "small corrections"
of order $\alpha/\pi$ as  in the LEP/SLC energy range; they are
essential parts of the dynamics.\par

A very important property of these large DL and SL logarithmic
terms is that their coefficients
reflect the gauge and Higgs structure
of the basic interactions. Thus, they offer a striking signature
for studying the underlying dynamics and differentiate between
the Standard (SM), and the Minimal Supersymmetric Standard
Model (MSSM), \cite{class, reality}.
Compared with the high level of the experimental accuracy
(few per mille), that is expected for future colliders,
this should allow  deep tests of the
basic electroweak interactions.

This has been illustrated recently by showing the relevance
of the SL and DL terms
at high energy colliders in the case of light and  heavy fermion
pair production in $e^+e^-$
\cite{log,MSSMlog} and $\gamma\gamma$ \cite{LR} collisions,
and also for sfermion pair production \cite{BMRV}.
Corrections including  higher order contributions have also
been computed \cite{BMRV, resum}.\par

Alternatively, these large logarithmic effects
may also appear  as large background contributions to possible
NP signals. It will therefore be essential to have  full control
on them, which necessitates
a  precise analysis of  the various virtual contributions
induced from each dynamical  sector. This will be particularly important
 in case a departure is  observed
 and hints for its origin are examined.  \par

The aim of this paper is to discuss these aspects in the
case of neutral gauge boson pair production in
$e^+e^-\to \gamma\gamma, ~Z \gamma, ~ZZ$. One loop effects
in SM have already been computed  some time ago
\cite{oldgg}. The additional step here is   to
analyze  in  detail  the high energy  behaviour,
and consider also  the complete set of MSSM contributions.
We examine how the asymptotic
(double log (DL) and single log (SL)) contributions are generated
using the complete expressions in each of the gauge and
Higgs sectors, in  SM and  MSSM.
This should also be useful  for
discussing possible modifications due to NP, and
proposing strategies for comparing with experimental
results.\par

Incidentally, the neutral gauge boson  production processes
$e^+e^-\to Z \gamma,~ZZ$, received recently considerable
theoretical \cite{NAGCt, NAGCt1, NAGCt2} and experimental \cite{NAGCe}
interest motivated by the search for anomalous neutral gauge boson
self couplings (NAGC). At tree level in SM and MSSM no couplings
exist among three neutral gauge bosons. Such couplings only appear
at one loop, through fermion triangles
involving leptons and quarks in SM, and also
charginos and neutralinos  in MSSM \cite{NAGCt1}.
These contributions are of course part of the complete
one loop corrections mentioned above.
The interest in them though is that additional
such  contributions may appear,  induced
from NP forms containing \eg heavier fermions, non perturbative structures,
or  even direct neutral boson couplings. To experimentally study such
couplings, it is  essential to have full control of the
"normal" SM or MSSM corrections;
we will devote a special discussion to this point.\par

Finally, we  consider the role of  longitudinal $Z$ production.
At high energy the $Z_L$ production  in SM and MSSM is strongly
depressed; but the depression is  stronger at the Born level,
than after radiative corrections are included. As a result,
the Born contribution to  \eg $Z_LZ_L$ production  above 1TeV,
is  negligible compared to the 1-loop one.
Such a channel may be
a suitable   place   to search for   forms of
anomalous NP contribution  generated by \eg
  a strongly interacting Higgs sector. \\

The paper is organized as follows. In Section 2 we describe the
method used for the computation of the one loop contributions
and we give the explicit expressions of the asymptotic amplitudes.
The numerical results are presented in Section 3 and we conclude
in Section 4. In Appendix A we define our notations and give
the expression of the Born terms.

\section{The one loop contributions}

In the present work, the complete 1-loop effects in
$e^-e^+ \to \gamma \gamma, ~ Z\gamma, ~ ZZ$
 have been computed, using  the on-shell renormalization
formalism \cite{Hollik}. The relevant diagrams
are depicted in Fig.1. The results are expressed in terms
of invariant amplitudes $N_j$ defined in Appendix A.
We refrain from giving the
explicit expressions of the invariant amplitudes in
this short paper, since they can
be found in the archives \cite{ourpaper}.
The structure of these amplitudes is
\bq
N_j(s,t,u) =N^{\rm ren+Born}_j+N^{\rm  Tri}_j+N_j^{\rm Box} ~ ~,
\label{Nj}
\eq
\noindent
where  $N^{\rm ren+Born}_j$ represent
the Born amplitudes modified by renormalization counter terms and
the gauge  and electron self-energy functions.
The other two terms in (\ref{Nj}) describe
the triangular and box diagram contributions.
The SM contribution is then fully determined
by diagrams involving only standard particle exchanges;
while in the MSSM case additional diagrams arise involving
SUSY contributions to self-energies and the triangle and box
diagrams containing chargino, neutralino, slepton, squark
and charged and neutral Higgs exchanges.
The couplings and masses are
taken from \cite{Rosiek}, and the neutralino
mixing formalism of \cite{LeMouel} is used.

In the illustrations we  use the specific benchmark MSSM models
corresponding to typical cases with either low or high masses
for the inos or the sfermions \cite{Ellis-bench, Snowmass}.
These models are constructed in the context of the constrained
MSSM SUGRA framework, with universal soft
supersymmetry-breaking masses, assuming
R-parity conservation. They have been selected to be consistent with
all present particle  physics and cosmological constraints, and
provide a rough indication of all available
possibilities in the SUGRA framework \cite{Ellis-bench}.

The 1-loop amplitudes are quite involved.
They become simple and intuitive though at
very high energies, whenever   $(s,~|t|,~ |u|)$  are   much
larger than all  masses of the particles exchanged
in the   various diagrams.
Under such conditions, the dominant radiative corrections are
described by single (SL)  and double-log  DL corrections
affecting only those $N_j$ which
receive non -vanishing Born-level
contributions; (these Born amplitudes are given in
( \ref{N3-N9-Born}- \ref{Nj-Born-Zg})).

 Thus, to the leading-log approximation,
the radiatively corrected invariant amplitudes  are given by
\bqa
N_j& \simeq &N^{ Born,L}_j~[1+c_L]
+N^{ Born,R}_j~[1+c_R] +d^{(W)}_{j,L}~,
\label{Nj-asym}
\eqa
where only the invariant amplitudes $N_j$ for which the
Born contribution is non-vanishing, appear.
The $c_{L,R}$ coefficients are process, model and $j$-independent.
In the \underline{ SM} case, they are given  by
\bqa
&& c_L=
{\alpha (1+2c^2_W)\over 16\pi s^2_Wc^2_W}
\Big [3 \ln{s\over \mwd}-\ln^2{s\over \mwd} \Big]
-{\alpha a^{(W)}\over8\pi s^2_W} \ln^2{s\over \mwd} ~~,
\label{cL-SMasym} \\
&& c_R=
{\alpha\over 4\pi c^2_W}~
\Big [3 \ln{s\over \mwd}-\ln^2{s\over \mwd}\Big ] ~~,
\label{cR-SMasym}
\eqa
while  in the  \underline{ MSSM} case by
\bqa
&& c_L=
{\alpha (1+2c^2_W) \over 16 \pi s^2_Wc^2_W}
\Big [2 \ln{s\over \mwd}-\ln^2{s\over \mwd}\Big ]
-{\alpha a^{(W)}\over 8\pi s^2_W} \ln^2{s\over \mwd} ~,
\label{cL-MSSMasym} \\
&&c_R=
{\alpha\over4\pi c^2_W}~
\Big [2 \ln{s\over \mwd}-\ln^2{s\over \mwd}\Big ]~~, \label{cR-MSSMasym}
\eqa
where in the right hand part of  (\ref{cL-SMasym}) and (\ref{cL-MSSMasym})
\bq
a^{(W)}=-2 ~~~~,~~~~ \frac{3-4s^2_W}{1-2s^2_W}~~~~,~~~~
\frac{4c^2_W}{1-2s^2_W}~~~, \label{aW} \\
\eq
should be used for $\gamma\gamma$, ~$\gamma Z$ and $ZZ$
respectively.

As seen from (\ref{Nj-asym}), $(c_L, ~c_R)$ provide
angular-independent universal  corrections to the Born amplitudes.
In addition to them though,  the additive
W-box  contributions $d_{j,L}^{(W)}$ appear in (\ref{Nj-asym}),
which are discussed below.\par

The $c_{L,R}$ correction in (\ref{Nj-asym})
 are (a least partly) induced  by vertex
diagrams involving photon, $Z$, $W$ and
MSSM partner exchanges.
In particular, the first terms for  $c_L$
and the complete $c_R$ terms
(compare (\ref{cL-SMasym}-\ref{cR-MSSMasym}) ), which have the
structure
\bq
\Big [3\ln \Big (\frac{s}{\mwd}\Big )-\ln^2\Big (\frac{s}{\mwd}\Big ) \Big ]
~~{\rm in ~~ SM}~~~,~~~
\Big [2\ln \Big (\frac{s}{\mwd}\Big )-\ln^2\Big (\frac{s}{\mwd}\Big ) \Big ]
~~{\rm in ~~ MSSM}~~,
\eq
 are generated by diagrams involving an
 electron line and  satisfy the same rules as those
established in \cite{reality,LR} for fermion and scalar pair
production in $e^-e^+$ annihilation. In agreement with
\cite{reality,LR},
we find that their  coefficients are independent of the
SM or MSSM scales, depending only on the $e_L, ~e_R$ weak
isospin $I_e$ and hypercharge $Y_e\equiv 2(Q_e-I_e^{(3)})$, through
\bq
\frac{\alpha}{4\pi} \Big [\frac{I_e(I_e+1)}{s^2_W}
+\frac{Y_e^2}{4c^2_W} \Big ] ~~ .
\eq
Thus, (as far as the electron-line  terms are concerned),
 at energies much larger than the masses of all standard and
supersymmetric particles, the only difference between SM and
MSSM dynamics
 is concentrated in changing the SL coefficient from 3 to 2, obviously
because of the different number of degrees of freedom in the two
models. On the contrary, the double-log term DL is the same in
both, the  SM and MSSM dynamics  \cite{reality,LR}.

The last $a^{(W)}$ terms in (\ref{cL-SMasym}, \ref{cL-MSSMasym}),
 come from sub-graphs involving the final gauge bosons, and only
 contain  DL terms. As in the electron-line case, the coefficients
 of  these  DL terms are insensitive to the differences
 between the  SM and MSSM dynamics; they are given in (\ref{aW}).

The  $d^{(W)}_{j,L}$ term in (\ref{Nj-asym}) is a specific
purely standard  W-box contribution, whose coefficient is fixed
by the $\gamma WW$ and  $ZWW$ couplings.
 It is angular dependent and of course,
 insensitive to  supersymmetric particle exchanges. It is  given by
\bqa
&& d^{(W)}_{j,L}= {\alpha^2b^{(W)} \over  s^2_W}P_L~\Big \{\eta^j_t
\Big [2 \ln{s\over m^2_W} \ln {1-\cos\theta\over 2} +
\ln^2 {1-\cos\theta\over 2}\Big ]\nonumber\\
&& +\eta^j_u~
\Big [2 \ln{s\over m^2_W}\ln{1+\cos\theta\over 2}
+\ln^2{1+\cos\theta\over 2}\Big ]\Big \} ~~, \label{d-asym-gg}
\eqa
\noindent
where again the index $j$ runs over the $N_j$ amplitudes that receive
non-vanishing Born contribution. For $\gamma\gamma$ production,
the parameters to be used in (\ref{d-asym-gg}) are
\bqa
&& b^{(W)}=1 ~~,  \nonumber \\
&& \eta^{1,2}_t={1\over t}~~,~~\eta^{4}_t={1\over t}+{2\over s}
~~, ~~\eta^{1,2}_u={1\over u}~~,~~\eta^{4}_u=-~{1\over u}-~{2\over s}
~~ ;
\eqa
while for $\gamma Z$
\bqa
b^{(W)}=1/4s_Wc_W  ~~ , & & \nonumber \\
\eta^{1,2}_t=-{s\over2}\eta^{6}_t={3-4s^2_W\over t}-{1\over u} &,&
\eta^{1,2}_u=-{s\over2}\eta^{6}_u={3-4s^2_W\over u}-{1\over t}
~, \nonumber \\
\eta^{4}_t={3-4s^2_W\over t}+{1\over u}+{8c^2_W\over s} &,&
\eta^{4}_u=-~{3-4s^2_W\over u}-{1\over t}-~{8c^2_W\over s}
~, \nonumber \\
\eta^{5}_t={1\over u}+{4c^2_W\over s} &,&
\eta^{5}_u=-~{3-4s^2_W\over u}-~{4c^2_W\over s} ~;
\label{d-asym-Zg}
\eqa
and for $ZZ$
\bqa
 b^{(W)}=1/2s^2_W ~~ , && \nonumber \\
\eta^{1,2}_t={1-2s^2_W\over t}-{1\over u} &,&
\eta^{1,2}_u={1-2s^2_W\over u}-{1\over t} ~, \nonumber \\
\eta^{4}_t={1-2s^2_W\over t}+{1\over u}+{4c^2_W\over s} &,&
\eta^{4}_u=-~{1-2s^2_W\over u}-{1\over t}-~{4c^2_W\over s}
 ~, \nonumber \\
\eta^{5}_t={1\over u}+{2c^2_W\over s} &,&
\eta^{5}_u=-~{1-2s^2_W\over u}-~{2c^2_W\over s}
 ~, \nonumber \\
\eta^{6}_t=-\eta^{8}_t=-~{2\over s}\Big [{1-2s^2_W\over t}
-~{1\over u}\Big ] &,&
\eta^{6}_u=-\eta^{8}_u=-~{2\over s}
\Big [{1-2s^2_W\over u}-~{1\over t}\Big ]  ~, \nonumber \\
\eta^{7}_t={2s^2_W-1\over t}-~{2c^2_W\over s} &,&
\eta^{7}_u={1\over t}+~{2c^2_W\over s} ~. \label{d-asym-ZZ}
\eqa

When  the helicity amplitudes generated by the
$N_j$ ones are computed, it is found  that the
 above structure of mainly
multiplicative corrections to
the Born contributions  is only preserved for the
TT  amplitudes, where both final gauge bosons are transverse.
It is only in this case that
the 1-loop asymptotic   amplitudes are given by
the Born ones, multiplied by the various leading-log  coefficients.
Such a factorization form
does not work for the TL, LT and  LL helicity amplitudes, which are
 mass suppressed; thereby forcing
  the difference between the high energy
behaviour of the 1-loop helicity amplitude, and its Born
contribution, to be  not  simply logarithmic, but to also have  a
power-law part.

As an example we recall that due to "gauge cancellations",
the Born $e^-e^+ \to ZZ$ TT amplitudes  behave like a constant
at asymptotic energies, the Born TL and LT ones vanish
like $m_Z/\sqrt{s}$, and the Born LL amplitudes diminish like
$m^2_Z/s$. This latter property  can
be explicitly seen in
\bqa
F^{Born}_{\lambda 00}& \simeq &-~(2\lambda){16m^2_Z\over
s}~ {\cos\theta\over\sin\theta} \Big \{{(2s^2_W-1)^2
\over4s^2_Wc^2_W}P_L+{s^2_W\over c^2_W}P_R\Big \}~~.
\label{born-asym-ZZ00}
\eqa

When the 1-loop effects are included, the TL and LT amplitudes
receive, apart from the  logarithmic factors,
additional mass dependent terms of the type
$M/\sqrt{s}$, where $M$ is some mass involved in the one loop diagrams.
For $Z_LZ_L$ amplitudes, these 1-loop modification  leads
to a strikingly  different high energy  structure, where  the rapidly
vanishing $\sim m^2_Z/s$ Born behaviour is replaced by a
logarithmically increasing one
involving $\ln^2{|t|/ M^2}$ and $\ln^2{|u|/ M^2}$ terms.
This structure is induced by  Higgs sector box
diagrams, whose asymptotic contribution dominates the
tree-level one.\par

The simplest way to obtain it,
is by using the equivalence theorem and considering the processes
$e^+e^-\to G^0 G^0$ ($G^0$ being the Goldstone state supplying
the longitudinal $Z_L$ component).
Since in the $m_e=0$ limit this later process has no Born term,
its only possible contribution comes from  boxes with internal $(eZHZ)$
and $(\nu WGW)$ lines; where $H$ stands for the
standard Higgs boson in SM, while  in
 MSSM  it represents a suitable mixture of the CP-even
 $H^0$ and the $h^0$ states.
 The resulting asymptotic helicity amplitudes is then
\bqa
&& F_{\lambda 00}
 \simeq
(2\lambda){\alpha^2 \sin\theta\over4}
\Big \{[\ln^2{|t|\over m^2_W}-\ln^2{|u|\over m^2_W}]\Big \}
\Big \{\Big ({1\over s^4_W}+{(2s^2_W-1)^2\over2s^4_Wc^4_W}\Big )P_L
+\Big ({2\over c^4_W}\Big )P_R\Big \} \nonumber \\
&& \simeq  (2\lambda){\alpha^2 \sin\theta\over 2}
 \ln \Big ({s\over \mwd}\Big )
  \ln\Big ({1-cos\theta\over 1+cos\theta}\Big )
\Big \{\Big ({1\over s^4_W}+{(2s^2_W-1)^2\over2s^4_Wc^4_W}\Big )P_L
+\Big ({2\over c^4_W}\Big )P_R\Big \}, \label{goldsm}
\eqa
in both, SM and MSSM.  Thus, at sufficiently  high energy,
the  order $\alpha^2$ contribution of (\ref{goldsm}),
becomes larger than the (suppressed) Born LL contribution
of (\ref{born-asym-ZZ00}).
The cross-over of these two terms
is around 1TeV.

Contrary to the TT case induced by  (\ref{Nj-asym}),
 asymptotically there is no
difference between the SM and the MSSM predictions for
longitudinal $Z_LZ_L$ production. This  is due to the fact
that the $H^0$ contribution is proportional to $\cos^2(\beta-\alpha)$
and the $h^0$ one proportional
to $\sin^2(\beta-\alpha)$, producing a result identical to
the SM one.\\

\noindent
{\bf The NAGC effects}\\
As mentioned in the Introduction, the full  1-loop results for
$e^-e^+\to Z\gamma, ~ZZ$ in SM or MSSM may be viewed as an
irreducible background in the search of possible
anomalous neutral gauge boson
couplings NAGC.  The general form of such couplings
has been written in \cite{NAGCt}. Here, we  restrict the  analysis
to the  on-shell couplings  $h^{\gamma, Z}_3$ and
$f^{\gamma, Z}_5$, which should be the dominant ones \cite{NAGCt1}.
Their contributions to the helicity amplitudes have been given
in \cite{NAGCt}. As shown there, $h^{\gamma, Z}_3$ contribute to the TT and
TL $e^+e^-\to Z \gamma $ amplitudes, and $f^{\gamma, Z}_5$ to the
TL $e^+e^-\to ZZ$ one. There is no contribution
to $e^+e^-\to \gamma\gamma$.\par

In \cite{NAGCt1}, dynamical models for generating NAGC couplings
have been considered. The conclusions of that work was that
the contributions to
$h^{\gamma, Z}_3$ and $f^{\gamma, Z}_5$ arising from 1-loop
effects induced by new higher fermions of mass $M$ pertaining to the NP
scale, diminish faster than $1/s$, for  $s \gg M^2$.
So they cannot  modify the leading-log SM or MSSM structure, and they are
always part of the subleading contributions.
But there may exist more general types of NAGC
leading to appreciable tree level
$h^{\gamma, Z}_3$ and $f^{\gamma, Z}_5$ couplings.

In any case, the processes  $e^-e^+\to Z\gamma, ~ZZ$ may  be used to
constrain such NAGC couplings, and for this purpose,  knowledge of the
complete 1-loop effects is essential.
Such constraints are presented in
the next Section, assuming  LC energies of  0.5 or  5 TeV.

\section{Numerical Results}

Here we  present  the numerical  prediction for  observables like
angular distributions, integrated cross sections and
asymmetries, defined in Appendix A.
Most of these observables do not refer to the final gauge
boson polarizations. But some remarks concerning the production
of specifically  transverse or longitudinal $Z$-bosons are also given.
\par

Due to the electron exchange diagrams in the $t$ and $u$ channels,
the angular distribution is strongly peaked in the forward and
backward directions. Because of detection difficulties
along the beam directions, we  only consider c.m.
scattering angles in the region $30^o< \theta < 150^o$.
The integrated  cross sections are thus  defined
by integrating in this angular region.\par

The  1-loop  radiative correction effects
to the differential and integrated cross sections
are described by the ratios to the corresponding Born
contributions.
Thus, the differential cross sections are described
in Figs.\ref{gg-differential-fig}-\ref{ZZ-differential-fig}
for SM and  a representative set of  MSSM SUGRA models
suggested in   \cite{Ellis-bench}, which are consistent
with all present particle and cosmological constraints.
The effects are always negative
and increase with  energy and the scattering angle.
In SM at 0.5 TeV, they are of about $-7\%$, $-8\%$ and $-15\%$
for $\gamma \gamma$, $Z\gamma $ and $ZZ$ production respectively;
while  at 5 TeV they correspondingly
 reach the level of  $-27\%$, $-40\%$, $-58\%$.\par

The differences between the SM and  the various
MSSM cases for the differential cross sections, especially
 at large angles,
 are within $1.5\%$ for $e^+e^-\to \gamma\gamma$, and
  increase to $10\%$ and  $20\%$ for
 $Z \gamma $ and $ZZ$ production
 respectively. \par

The radiative corrections effects to  the integrated
\underline{unpolarized} (summed over all final polarization)
cross sections are  described in
Figs.\ref{gg-sig-fig}-\ref{ZZ-sig-fig}.
The above 1TeV behaviour  of these cross sections
agrees  (apart from a model dependent constant term)
with the asymptotic leading  log expressions
 (\ref{cL-SMasym}-\ref{cR-MSSMasym}), for SM and MSSM
respectively. As pointed out in Section 2,
 the main difference between the SM and MSSM predictions
at high energy, stems from the respective factors
$(3\ln s-\ln^2 s)$ and $(2\ln s-\ln^2 s)$,
which depend only on the overall structure of the theory and are totally
independent of any other MSSM parameter. So, an
experimental measurement of the coefficient of
the linear log term, could check the agreement with SM or with
MSSM, and even provide hints if any additional NP contribution
is needed.\par

Another interesting quantity is the Left-Right polarization asymmetry
$A_{LR}$, which is not  affected by normalization
uncertainties, and its measurement may be
extremely interesting experimentally.
In this case the Born contribution is constant in energy
and satisfies
$A_{LR}(Born)=0,~0.2181,~0.4164$ for $\gamma\gamma,~\gamma Z,~ZZ$
respectively. The radiative corrections can then conveniently described
by the 1-loop induced departure from  these values. Since the effects are
similar to those for the integrated cross sections, we
do  not present them explicitly.

The above behaviour of the unpolarized cross section (or $A_{LR}$ asymmetry)
is ensured by the dominance of the TT amplitudes. In $\gamma\gamma$,
where only such amplitudes occur, we have  checked that by putting
an additional constant to the expression for  the asymptotic
cross section fitted at 5 TeV, we  get agreement with the
exact 1-loop result at the permille
level, for energies as low as  0.2 TeV.

In $\gamma Z$ production, the presence of TL amplitudes,
which are negligible in the several
TeV range\footnote{The TL cross section behaves
like  $M^2/s^2$ at high energies, whereas the TT one behaves
like $1/s$, apart from log factors.},
generate corrections below 1 TeV which are at the
several percent level and sensitive to the MSSM model considered.

The case of $ZZ$ production is more
interesting, because, in addition to the TL and LT components
which behave like the ones in $\gamma Z$, there are also LL
components. The Born LL component, which contributes
 about $10\%$  close to  threshold, is  strongly
depressed at higher energies, behaving like
$\sigma_{Born}^{LL} \sim 1/s^3$. But because of
large contributions generated by the Higgs
sector\footnote{Compare  the
Goldstone contributions given in (\ref{goldsm}). },
a logarithmic increase  arises above 1 TeV, illustrated
in Fig.\ref{ZZLL-sig-fig}.
This happens at a level which is  hardly observable,
except with very high luminosity colliders. Nevertheless we
show it, because of its exceptional behaviour
in the very high energy range. Its dependence on the Higgs mass is
rather weak; the relative differences between  the cross
sections for    $m_H=0.3$ TeV  or  $m_H=1$ TeV,
and the one for  $m_H=0.113$ TeV  being at the
permille level. \\

\vspace{1cm}

\noindent
{\bf Constraints on NAGC contributions}\\
We  first look at the "normal" NAGC contributions arising
from fermionic triangular loops, both in SM and  MSSM.
Since  these
decrease faster than $1/s$  with energy \cite{NAGCt1},
one can only look for NAGC  at energies  not higher than 1 TeV.
In $e^+e^-\to Z \gamma $ the effects are due to the  TT and TL
amplitudes; while  in $e^+e^-\to ZZ$ only  TL (and LT) amplitudes
contribute.
However, in the $Z \gamma $ case there is no interference between the
NAGC and Born TT amplitudes; since the final gauge boson helicities
are equal for the  NAGC amplitudes, and opposite for the Born ones.
So in both the $\gamma Z$
and $ZZ$ cases, the effects will mainly
come from the interference with the weaker Born
LT amplitudes, and should  be at the permille
level around (or below) 1 TeV.
We conclude therefore that  new NAGC contributions
generated  by \eg triangles
involving higher mass fermions,  will  be marginally
observable;  unless  very high luminosities  are available.

We  next look  at the sensitivity to the "true" NAGC amplitudes,
described in a model independent way by the phenomenological
coupling constants  $h^{\gamma, Z}_3$ for  $Z \gamma $,
and $f^{\gamma, Z}_5$ for  $ZZ$ \cite{NAGCt}.
Assuming a given experimental accuracy (for example a
conservative $1\%$)
on the unpolarized integrated cross sections $\sigma_{unp}$
and the Left-Right asymmetry $A_{LR}$, we then obtain the
  NAGC observability limits for such contributions.
This is illustrated in  Figs.9a,b for
$Z \gamma $ and $ZZ$ production, at  energies of 0.5TeV and 1TeV.

Note from Figs.9a,b,  that the $\sigma_{unp}$ and
$A_{LR}$ constraints are almost
orthogonal, allowing a good limitation of both the
 photon- and  $Z$-NAGC couplings. This arises because, in $\gamma Z$,
$\sigma_{unp}$ mainly depends on  $h^{\gamma}_3$,
whereas $A_{LR}$ is more  sensitive to $h^{Z}_3$. In $ZZ$
the roles of the photon and  $Z$ are interchanged,
$\sigma_{unp}$ mainly depending  on  $f^{Z}_5$, and
 $A_{LR}$ on  $f^{\gamma}_5$,
due to the different chirality structure of the
Born terms. All this can be traced off from the fact that
the photon couples vectorialy, while the  $Zee$ coupling
is essentially purely axial.

As seen from Figs.9a,b,
the implied sensitivity is likely to increase with energy.
We conclude therefore
that NP forms inducing the NAGC couplings
$h^{\gamma, Z}_3$, $f^{\gamma, Z}_5$ at the level
of $10^{-3}$ to $10^{-4}$, should be observable
around 1 TeV.

\section{Physics issues and Conclusions}

In this paper we have analyzed the behaviour of the
electroweak corrections to the processes
$e^+e^-\to\gamma\gamma, ~Z \gamma, ~ZZ$ at the one loop level,
in the context of the SM and the MSSM.\par

These processes are particularly interesting in various aspects.
Their final gauge bosons are easy to detect experimentally,
while their theoretical structure  provides clean tests
of the  electroweak interactions.
The Born terms are only due to electron exchanges in the
$t$ and $u$ channels; contrary to the $WW$ case, there is no
s-channel tree level term.  Since there are no QCD or  Yukawa
contributions,
the identification of the electroweak corrections
should  be very clean. \par

We have completely computed the 1-loop SM and MSSM corrections,  in
order to analyze the contents of the
 gauge, Higgs and Goldstone sectors, as well
as their supersymmetric counterparts. We have studied  how
these contributions vary  with energy, and how they
conspire  to generate asymptotically the leading  logarithmic
terms,  and in particular
the remarkable difference between the SM combination
$(3\ln s-\ln^2 s)$ and the MSSM one  $(2\ln s-\ln^2 s)$.
This contribution fixes the asymptotic form of
the transverse-transverse final gauge boson (TT)  amplitudes.
At moderate energies, we have also
found that the mass-suppressed TL and LL amplitudes play a non negligible
role.\par

The structures of the  angular distributions,
integrated cross sections and Left-Right asymmetries, for both
unpolarized and polarized final states, have also been studied numerically.
It was found in particular that the $A_{LR}$ asymmetry shows essentially
the same effects as the unpolarized cross section,
a feature which may be experimentally
interesting in reducing the normalization uncertainties.\par

In addition to studying the standard and  supersymmetric
effects,
we have considered possible additional NP NAGC contributions,
 as described by Effective Lagrangians. We have shown that
the corresponding coupling constants could be constrained
at an interesting level.\par

Summarizing, we can say that the electroweak radiative
corrections  are large and growing
with  energy. Starting from a few percent at the
energy range of a few hundreds of GeV, they reach  $10\%$ already at
1 TeV, and continue  growing according to the logarithmic
rules. So, in the high energy range they are no more
"small corrections". They are essential parts of the dynamics,
which can be experimentally analyzed at the
future colliders, whose accuracy
should reach the percent level or even better. For more accurate
theoretical predictions, computations of higher orders
may be attempted; it has already been claimed that
several logarithmic terms  exponentiate \cite{resum},
so that all the features that we have observed at the one loop
level,  remain true for  higher orders also.\par

We have also shown that  measurements of the three processes
$e^+e^-\to\gamma\gamma,~\gamma Z$ and $ZZ$ should provide
global tests of the basic interactions.
A strategy for these tests could be the following.
One could first try to compare the high energy behaviour of
the cross sections and Left-Right asymmetries with
the SM predictions, (containing in particular the
$(3\ln s-\ln^2 s)$ term). If there is no new particle produced,
SM may seem a reasonable assumption, and one would either
check its consistency or, if the check
fails\footnote{\ie  A disagreement with the predicted
high energy behaviour is identified.},
we  would be led to anticipate  some form
of NP.

Another situation may be that
many candidates for supersymmetric particles are found,
with masses considerably
smaller than the highest collider energy attainable.
One should then compare the high energy behaviour
with the MSSM prediction (containing in particular the
$(2\ln s-\ln^2 s)$ term) and again we could check whether
a global agreement appears.
If there are still some departures indicating the need
for additional NP effects, a comparison of the three processes
$\gamma\gamma$,  $Z \gamma $, $ZZ$ may provide a
hint for their origin; particularly  if
NP is related to NAGC couplings or to some other  anomalous
properties of the $Z$ boson.
We also note that  a measurement of the coefficient of the
double log term could  check if there are no higher gauge bosons
acting.\par
In conclusion the three processes studied here, present a large
panel of interesting properties. They are extremely
simple at Born level, but extremely rich in the supplied information
 at the radiative corrections  level.
The $\gamma\gamma$, $Z\gamma$ and $ZZ$ final states
are complementary for the study of the gauge (gaugino) sector in
the MSSM models, the Higgs sector and the search for
Neutral Anomalous Gauge Couplings.\par

We have thus seen one more aspect of the fact that in
the several TeV domain the electroweak interactions start
becoming strong.
The processes studied here illustrate this property and
should be considered as part of the research program
at the future high energy colliders,  demanding for
the highest luminosities. In some respect, the
  tests supplied by $(e^-e^+\to \gamma \gamma, ~ Z\gamma, ~ ZZ)$,
     are in the same spirit
as the high precision tests performed with $g-2$
measurements, or with $Z$ peak physics. They should provide global
checks of the validity of the SM or MSSM theory.\par

\newpage

\renewcommand{\theequation}{A.\arabic{equation}}
\renewcommand{\thesection}{A.\arabic{section}}
\setcounter{equation}{0}
\setcounter{section}{0}

{\large \bf Appendix A: Kinematical details.}\\

The  considered  process is
\bq
e^-(\lambda, l)~+~e^+(\lambda', l')
\to V(e,p)+ V'(e',p') ~~ , \label{process}
\eq
where  $(l,~l')$ are the  incoming electron and positron momenta,
and $(\lambda,~\lambda')$ their  helicities.
The outgoing neutral gauge bosons $Z$ or $\gamma$ are generally
denoted as $V$ and $V'$, their momenta  as   $(p,~p')$ respectively,
the complex conjugate of their polarization vectors
as $(e, ~ e')$, and their corresponding  helicities as
$(\mu, \mu')$. We also define
\bqa
&& q=l-p=p'-l' ~~~~ ,~~~~  q'=l-p'=p-l' ~~
\nonumber \\
&& s=(l+l')^2=(p+p')^2~~,~~t=q^2~~,~ ~
u=q'^2~ .
\nonumber
\eqa
The c.m. scattering angle between $\vec l$ and $\vec p$ is denoted
by $\theta$.\par

The electron mass is throughout  neglected, implying
$\lambda'= - \lambda= \pm 1/2 $. Consequently there are at most
18 helicity amplitudes  written as
\bqa
F_{\lambda,\mu,\mu'}&\equiv&
F[ e^-( \lambda , l )~ e^+(\lambda'=-\lambda,
 ~ l')~ \to ~V (e(\mu),p) ~V'(e'(\mu'),p') ]\nonumber\\
&&=\sum_{j=1,9} \bar{v}(\lambda', l')~I_j~N_j(s,t,u, \lambda)
~u(\lambda, l)~~, \label{helicity-amplitudes}
\eqa
in terms of  nine  Lorentz invariant forms
 $I_j$, ~($j=1,9$) defined below.
\bqa
I_1=(e \cdot l)(\gamma \cdot e') &,~~
I_2=(e' \cdot l)(\gamma \cdot e) & ,
~~ I_3=(e\cdot l)(e'\cdot l)(\gamma\cdot p),~~
\nonumber \\
I_4=(e\cdot e')(\gamma \cdot p) & , ~~
I_5=(e \cdot p')(\gamma \cdot e') & ,~~
I_6=(e \cdot p')(e'\cdot l)(\gamma \cdot p)~~,
\nonumber \\
I_7=(e' \cdot p)(\gamma \cdot  e) &,~~ I_8=(e'\cdot p)(e\cdot l)
(\gamma \cdot p)&,
~~ I_9=(e \cdot p')(e'\cdot p)(\gamma \cdot p) ~. \label{Ij}
\eqa
Their coefficients,  split according to the electron-helicity,
define  the invariant amplitudes as
\bq
N_j(s,t,u,\lambda )
 \equiv N^L_j(s,t,u)P_L+N^R_j(s,t,u)P_R ~~, \label{NjLR}
\eq
\noindent
where
\bq
P_L=\frac{1}{2} -\lambda ~~~,~~~P_R=\frac{1}{2} +\lambda ~~.
\label{proj}
\eq \\

In the specific case of   the process
\underline{$e^-e^+\to\gamma\gamma$},
only 4  transverse-transverse amplitudes appear,
which are described through the invariant functions
$N_1$, $N_2$, $ N_3$, $N_4$.\par

In the case of   \underline{$e^-e^+\to Z \gamma $},
where the gauge boson polarization and momenta are
defined by  $Z(e,p)$ and $\gamma(e',p')$,
the process is most generally described by the 6 invariant amplitudes
$N_1, ...~ N_6$.\par

Finally,  for \underline{$e^-e^+\to ZZ$}, the complete set of
 the $N_1, ... ~N_9$  amplitudes is needed for a complete description.\\

\noindent
{\bf Observables}\\
The polarized differential cross sections is written as
\bq
{d\sigma(\lambda,\mu,\mu')\over d\cos\theta}
={\beta\over 32 \pi s}~C_{stat}
~|F_{\lambda,\mu,\mu'}|^2~~, \label{polarized-dsigma}
\eq
\noindent
where  $C_{stat}=1/2,~1/2,~1$ for $\gamma\gamma,~ZZ,~\gamma Z$,
respectively. The integrated cross sections for definite polarizations
are
\bq
\sigma(\lambda,\mu,\mu')=\int^{c}_{-c} d\cos\theta~~
{d\sigma(\lambda,\mu,\mu')\over d\cos\theta} ~~,
\eq
\noindent
where $c\equiv \cos\theta_{min}$ is an angular cut (fixed
at $\theta_{min}=30^0$ in the numerical applications).

For longitudinally polarized $e^{\pm}$ beams, the Left-Right
polarization asymmetry is defined as
\bq
A_{LR}(\mu,\mu')=
\frac{\sigma(-{1\over2},\mu,\mu')-\sigma(+{1\over2},\mu,\mu')}
{\sigma(-{1\over2},\mu,\mu')+\sigma(+{1\over2},\mu,\mu')} ~~~.
 \label{ALR-definition}
\eq \\

\noindent
{\bf The Born terms}

These  are due to electron exchange in the $t$ and $u$ channels.
In terms of the invariant amplitudes  defined in (\ref{NjLR}),
they are written as
\bq
N_j^{\rm Born} = N_j^{\rm Born,~t} +N_j^{\rm Born,~u} ~~,
\label{Nj-Born}
\eq
and for all processes it is found that
\bq
N^{\rm Born,~t}_{3,9}=N^{\rm Born,~u}_{3,9} =0 ~~ .
\label{N3-N9-Born}
\eq
The rest of the invariant amplitudes  are:

$\bullet$~ \underline{$e^-e^+\to\gamma\gamma$.}
\bqa
&& N^{\rm Born,~t}_1=N^{\rm Born,~t}_2=N^{\rm  Born,~t}_4
=  -~\frac{e^2}{t}P_L
-~\frac{e^2}{t}P_R
 ~~ , \nonumber \\
&& N^{\rm Born,~u}_1=N^{\rm Born,~ u}_2=
- N^{\rm Born,~ u}_4
=  -~\frac{e^2}{u}P_L
-~\frac{ e^2}{u}P_R
 ~~. \label{Nj-Born-gg}
\eqa

\noindent
$\bullet$~ \underline{$e^-e^+\to Z Z$.}
\bqa
&& N^{\rm Born,~t}_1
=  N^{\rm Born,~t}_2=N^{\rm Born,~t}_4
=-\, \frac{s}{t-\mzd} \, N^{\rm Born,~t}_5
  = -\, \frac{s}{2}\, N^{\rm Born,~t}_6\nonumber \\
&&
=-\, \frac{s}{s-t+\mzd }\, N^{\rm Born,~t}_7=
 \, \frac{s}{2}\, N^{\rm Born,~t}_8=
 -\, \frac{g^2_{ZL}}{t}\,-
 \, \frac{g^2_{ZR}}{t}\, ,
 \nonumber \\
&& N^{\rm Born,~u}_1
= N^{\rm Born,~u}_2=-N^{\rm Born,~u}_4
= \frac{-s}{s-u +\mzd} \, N^{\rm Born,~u}_5
\nonumber \\
&& =  \frac{-s}{2}\, N^{\rm Born,~u}_6=
\, \frac{-s}{u -\mzd }\,  N^{\rm Born,~u}_7=
   \frac{s}{2}~ N^{\rm Born,~u}_8=
 -\, \frac{g^2_{ZL}}{u}\, -\,
 \frac{g^2_{ZR}}{u}~  ,
\label{Nj-Born-ZZ}
\eqa
\noindent
with
\bq
g_{ZL}=e ~{(2s^2_{W}-1)\over2s_{W}c_{W}}~~~~~, ~~~~~~
g_{ZR}=e ~{s_{W}\over c_{W}} ~~.
\label{gZLR}
\eq
\noindent
$\bullet$~ \underline{$e^-e^+\to Z \gamma $.}
\bqa
&&\frac{s}{s-\mzd} ~ N^{\rm Born,~ t}_1
= \frac{s}{s +\mzd}~ N^{\rm Born,~ t}_2= N^{\rm Born,~t}_4
\nonumber \\
&&
=-~ \frac{s}{t-\mzd}~ N^{\rm Born,~t}_5
= - ~ \frac{s}{2}~ N^{\rm Born,~t}_6=
 \frac{eg_{ZL}}{t}P_L
+~ \frac{e g_{ZR}}{t}P_R ~~ ,
 \nonumber \\
&&
\left (\frac{s}{s-\mzd}\right ) N^{\rm Born,~u}_1
= \left (\frac{s}{s +\mzd}\right ) N^{\rm Born,~u}_2
= -  N^{Born,~u}_4\nonumber \\
&&
=  \left (\frac{-s}{s -u}\right ) N^{\rm Born,~u}_5
 = \frac{-s}{2}~ N^{\rm Born,~u}_6=
 \frac{eg_{ZL}}{u}P_L ~ +~\frac{ e g_{ZR}}{u}P_R ~~ .
 \label{Nj-Born-Zg}
\eqa

\clearpage
\newpage

\begin{figure}[th]
\[
\epsfig{file=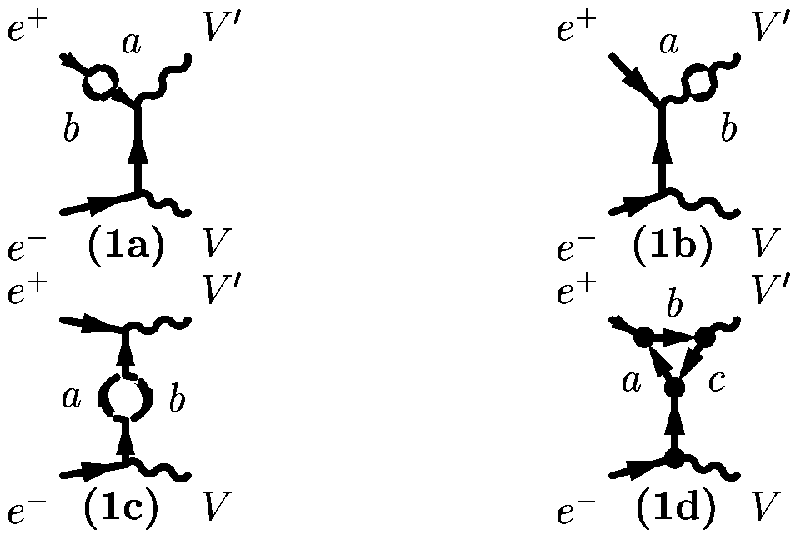,height=8.5cm,width=13cm}
\]
\[
\epsfig{file=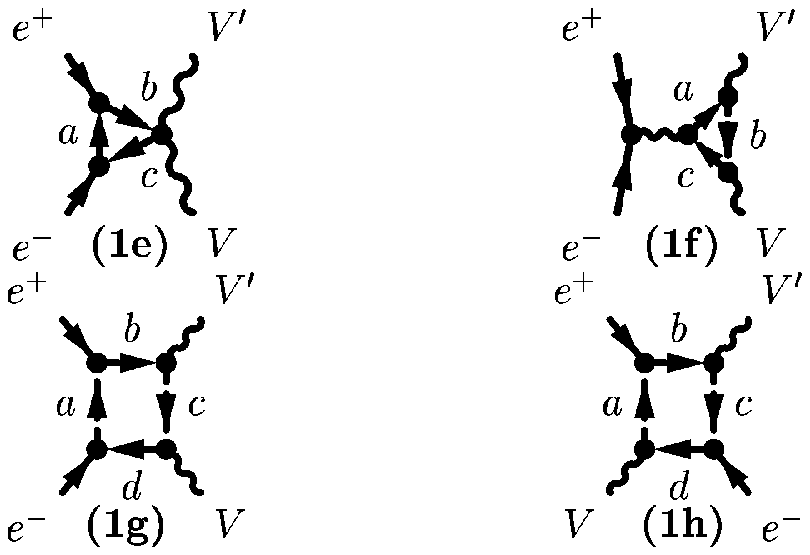,height=8.5cm,width=13cm}
\]
\caption[1]{Diagrams at one loop}
\end{figure}

\clearpage
\newpage

\begin{figure}[p]
\vspace*{-3cm}
\[
\hspace{-0.5cm}\epsfig{file=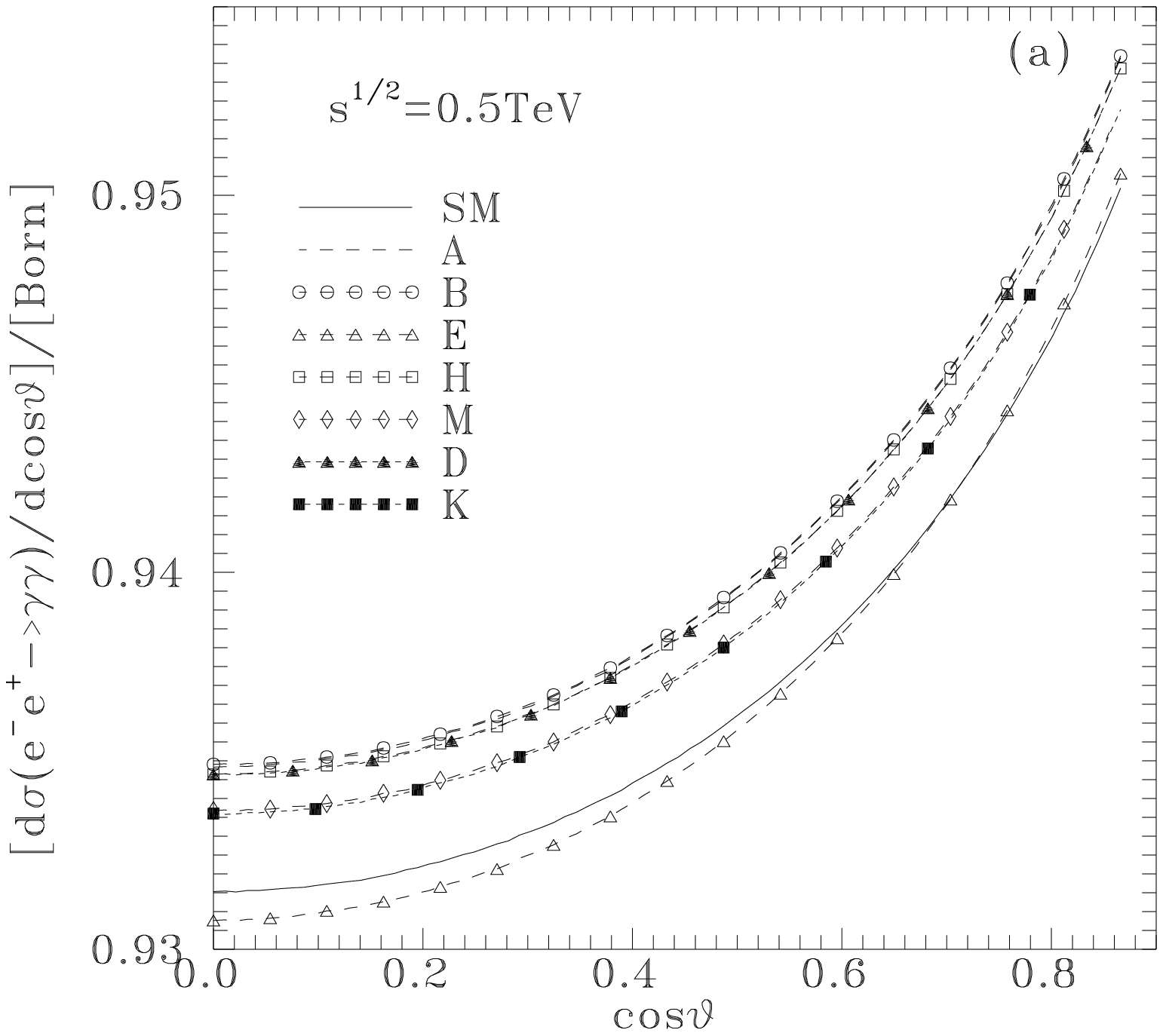,height=7cm}\hspace{0.5cm}
\epsfig{file=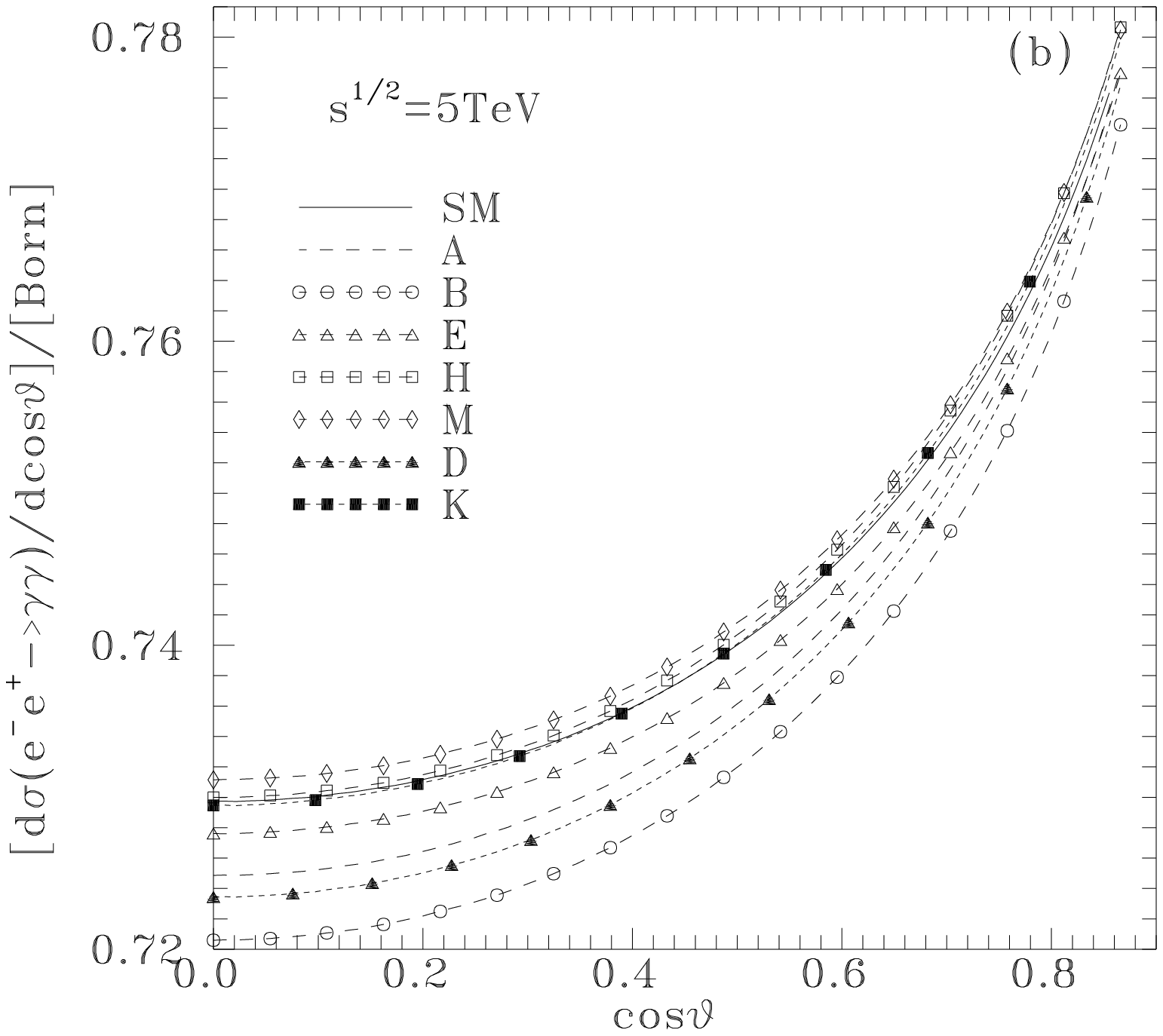,height=7cm}
\]
\vspace*{-0.5cm}
\caption[1]{The ratio of the unpolarized differential
cross section  $e^-e^+ \to \gamma \gamma $ to the Born
contributions, at 0.5TeV (a) and 5TeV (b), for  SM and
a representative subset
of the benchmark MSSM models of \cite{Ellis-bench}.}
\label{gg-differential-fig}
\end{figure}

\newpage

\begin{figure}[p]
\vspace*{-3cm}
\[
\hspace{-0.5cm}\epsfig{file=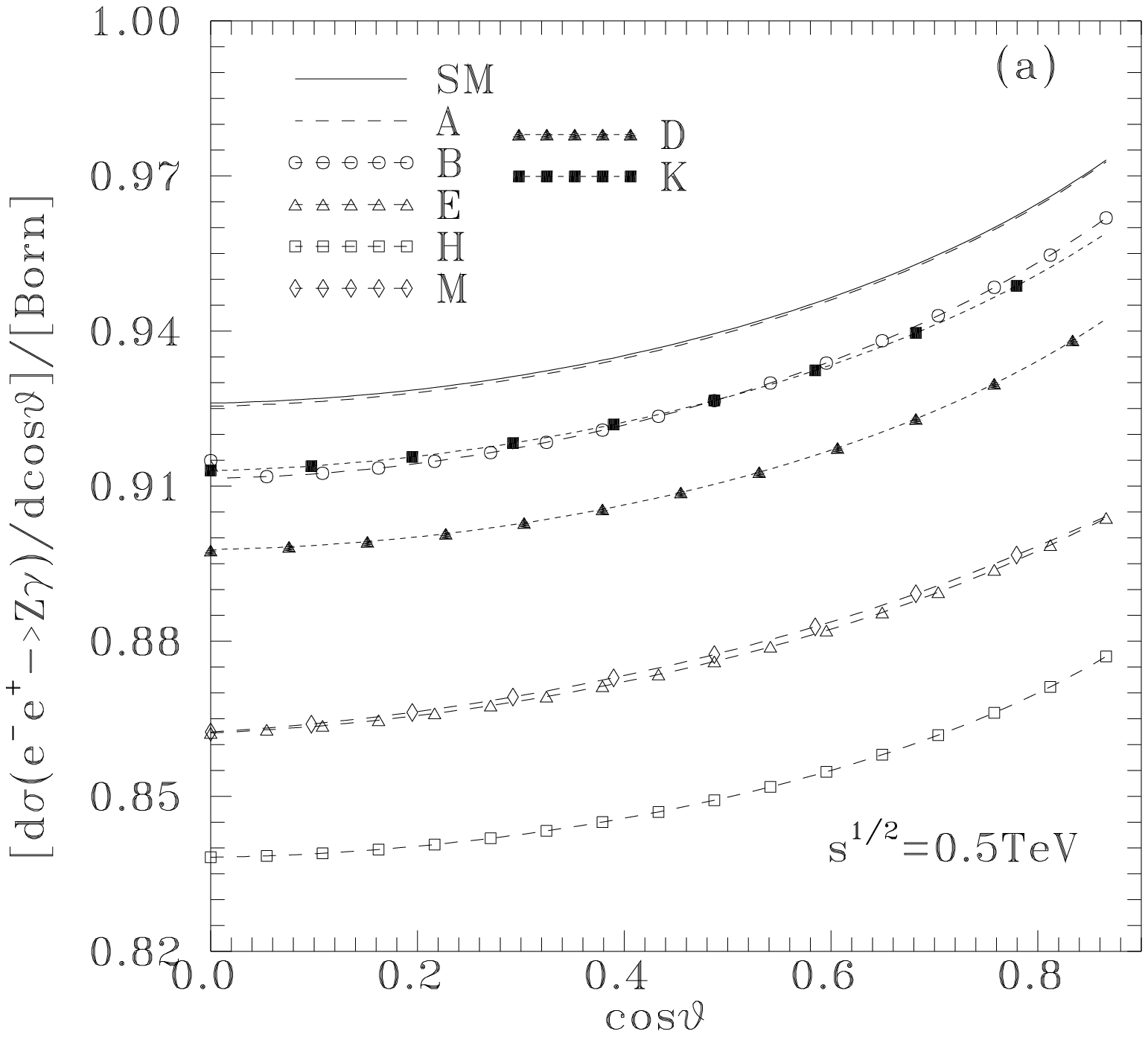,height=7cm}\hspace{0.5cm}
\epsfig{file=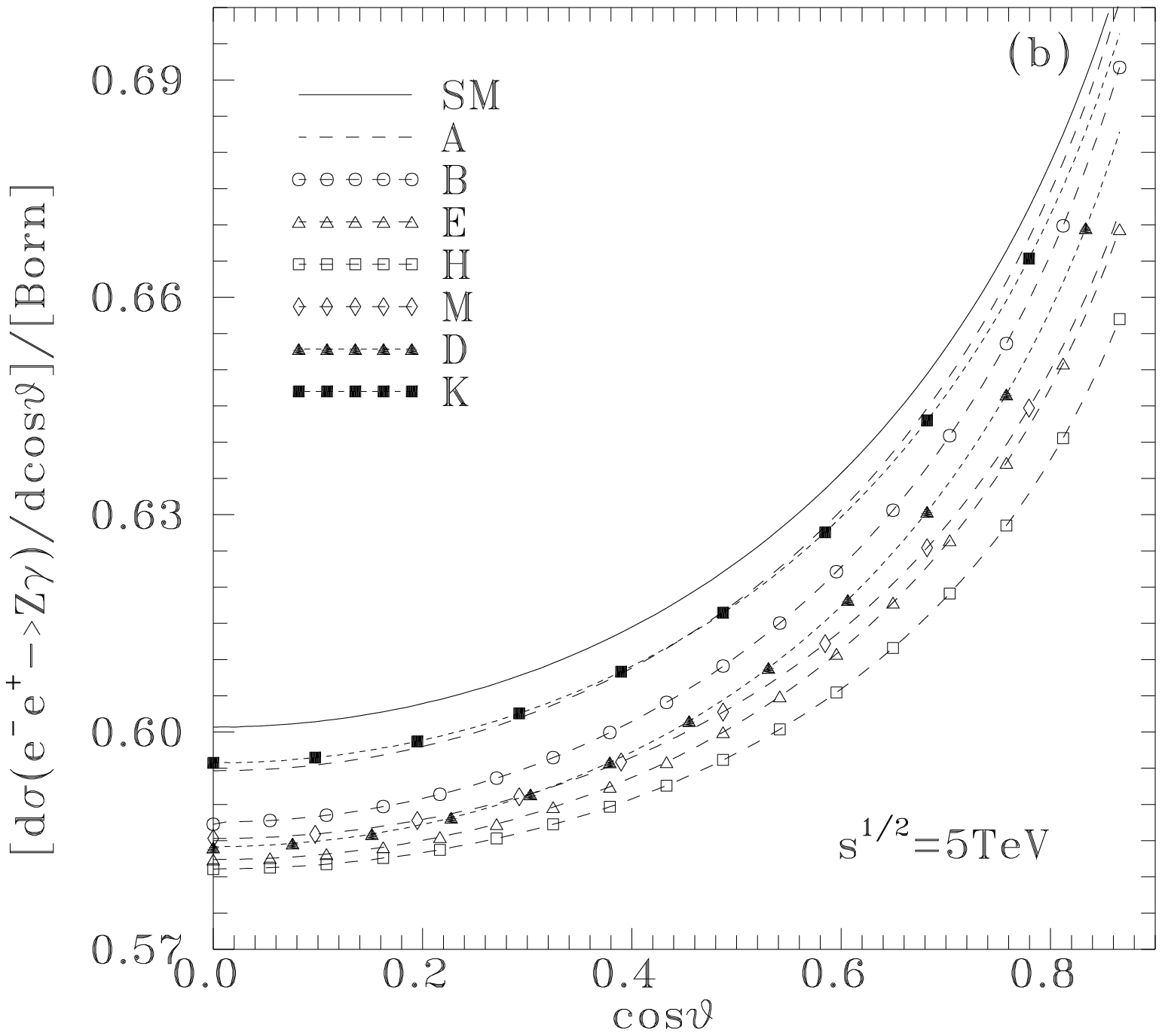,height=7cm}
\]
\vspace*{-0.5cm}
\caption[1]{The ratio of the unpolarized differential
cross section  $e^-e^+ \to Z \gamma  $ to the Born contribution,
at 0.5TeV (a) and 5TeV (b), for  SM and
a representative subset
of the benchmark MSSM models of \cite{Ellis-bench}.}
\label{Zg-differential-fig}
\end{figure}

\clearpage
\newpage

\begin{figure}[p]
\vspace*{-3cm}
\[
\hspace{-0.5cm}\epsfig{file=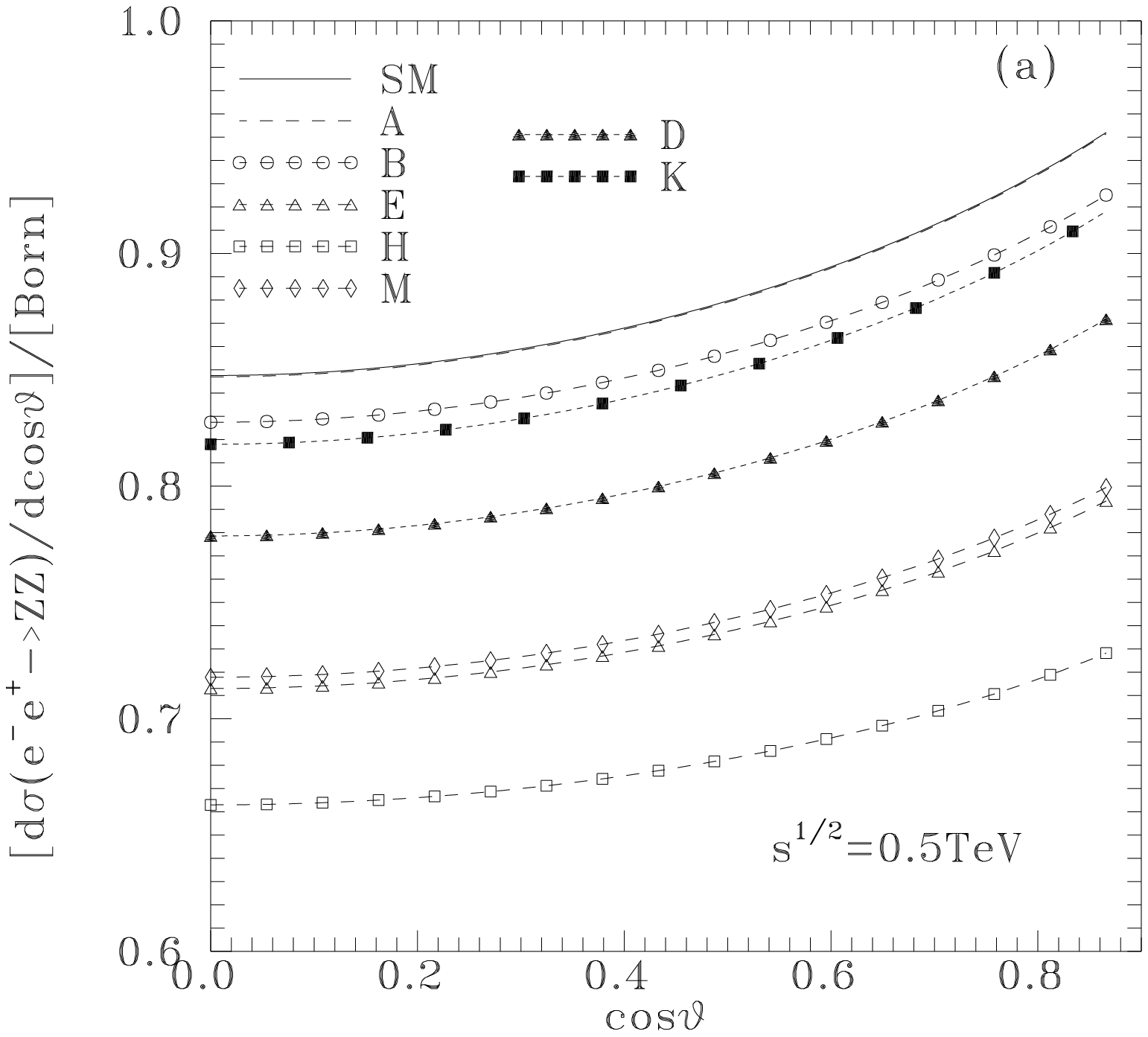,height=7cm}\hspace{0.5cm}
\epsfig{file=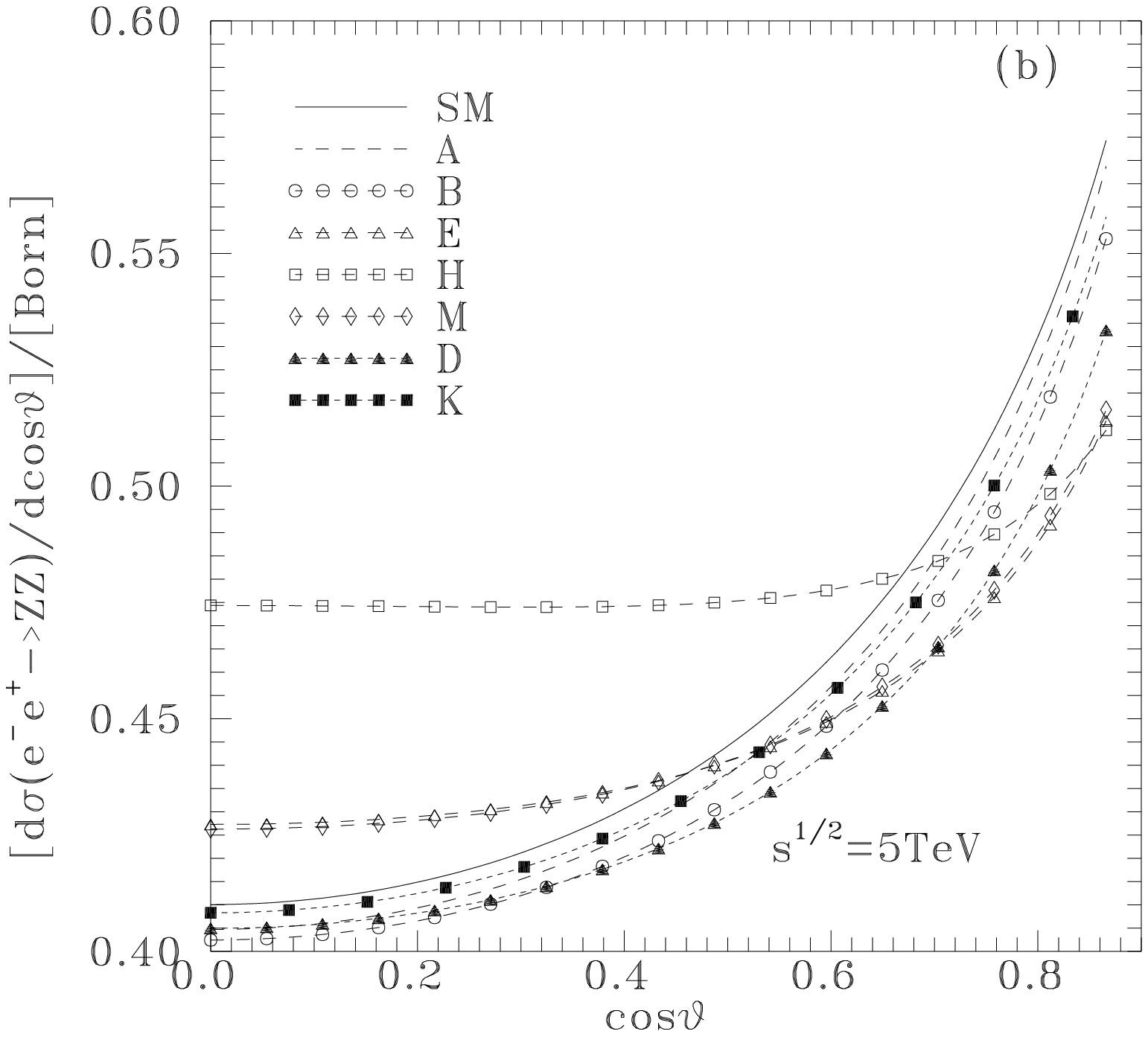,height=7cm}
\]
\vspace*{-0.5cm}
\caption[1]
{The ratio of the unpolarized differential
cross section  $e^-e^+ \to Z Z $ to the Born contribution,
at 0.5TeV (a) and 5TeV (b), for  SM and
a representative subset
of the benchmark MSSM models of \cite{Ellis-bench}.}
\label{ZZ-differential-fig}
\end{figure}

\clearpage
\newpage

\begin{figure}[b]
\vspace*{-2cm}
\[
\epsfig{file=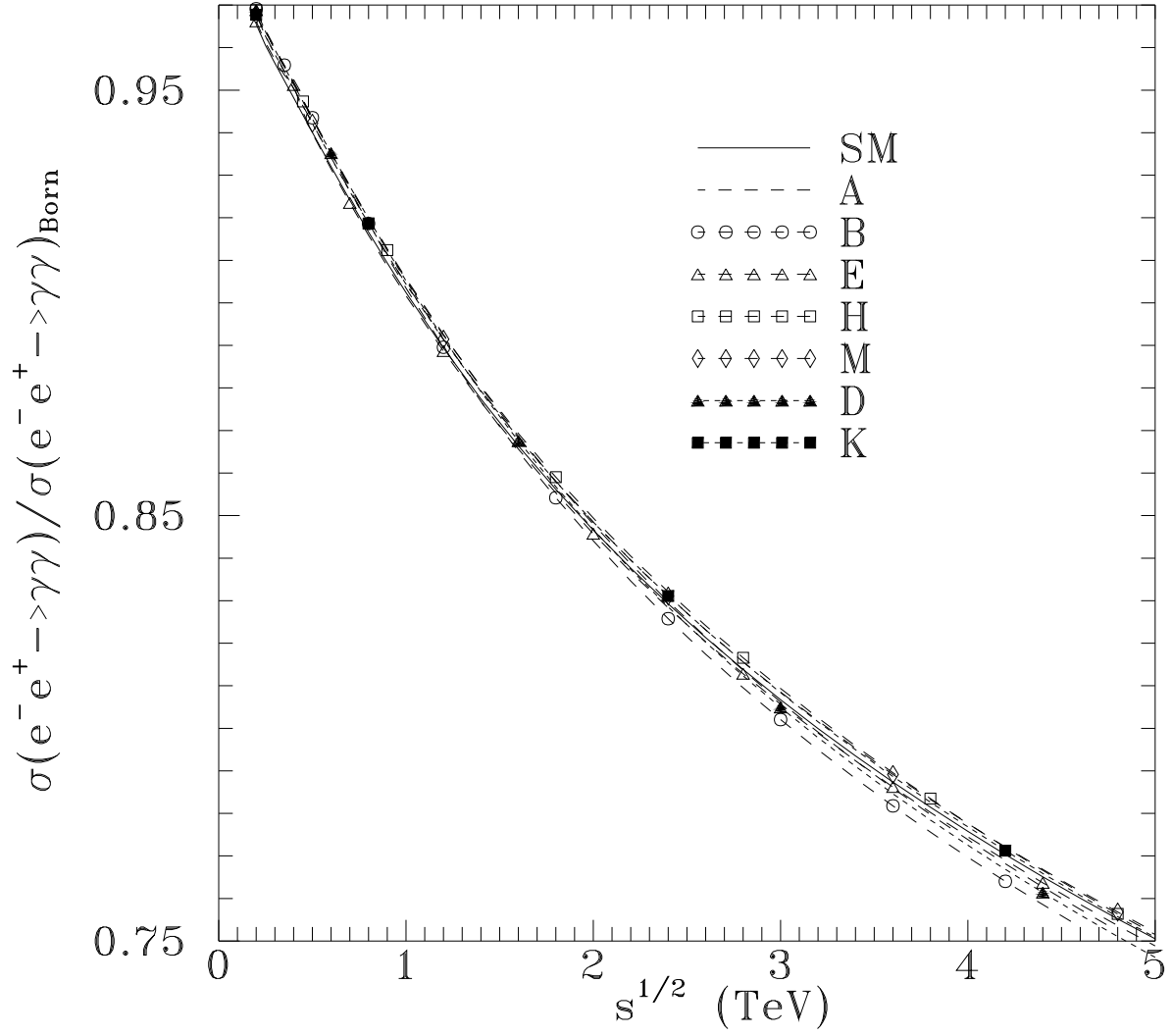,height=14cm}
\]
\vspace*{-0.5cm}
\caption[1]{The ratio of the unpolarized integrated
$\sigma(e^-e^+\to \gamma \gamma)$ cross section
to the Born contribution  for  SM and a representative subset
of the benchmark MSSM models of \cite{Ellis-bench}.}
 \label{gg-sig-fig}
\end{figure}

\clearpage
\newpage

\begin{figure}[p]
\vspace*{-2cm}
\[
\hspace{-0.5cm}\epsfig{file=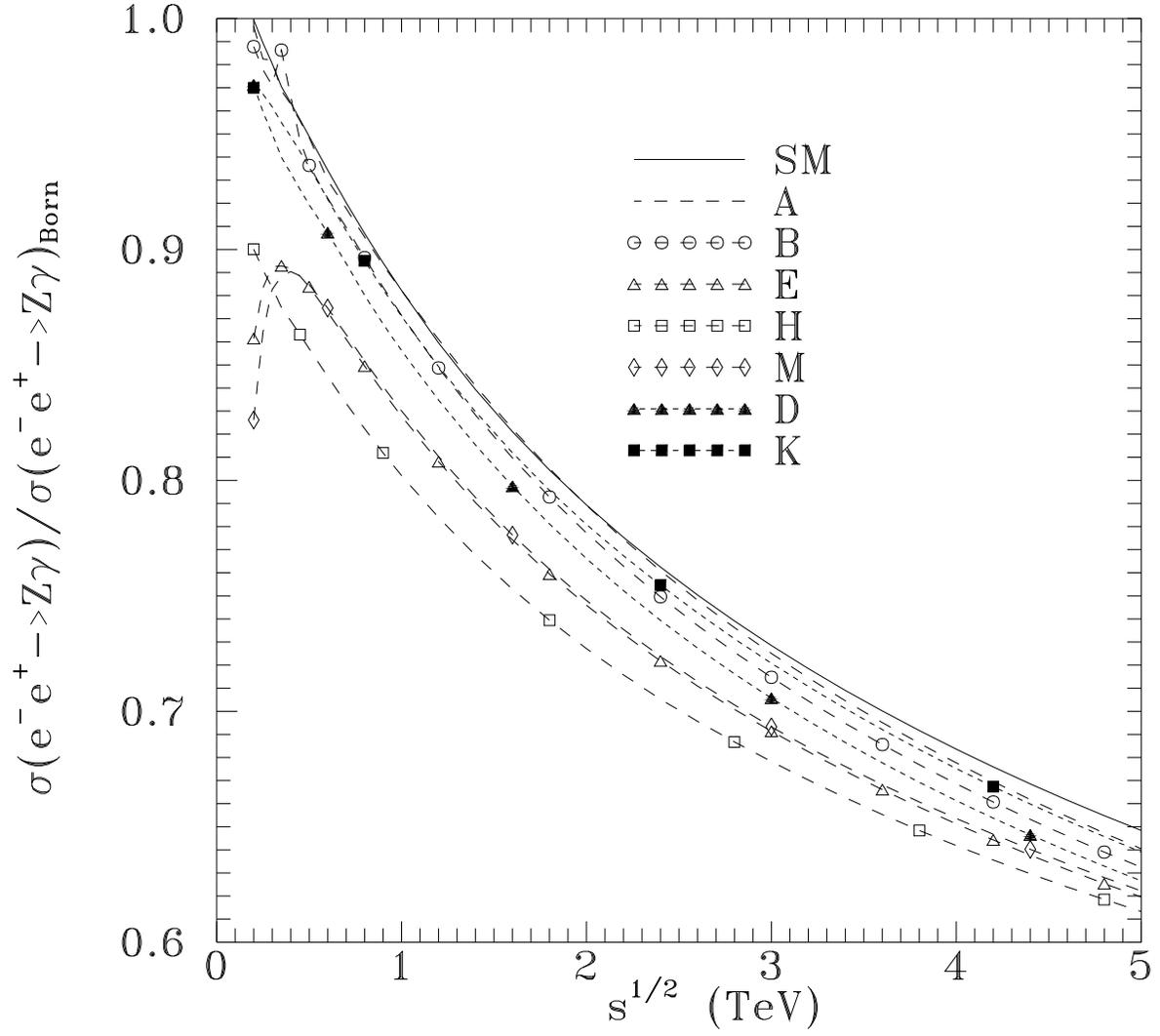,height=14cm}
\]
\vspace*{-0.5cm}
\caption[1]{The ratio of  the integrated unpolarized
$\sigma(e^-e^+\to Z \gamma )$ cross section to the Born contribution,
  for SM and a set of  MSSM models of \cite{Ellis-bench}.}
\label{Zg-sig-fig}
\end{figure}

\clearpage
\newpage

\begin{figure}[p]
\vspace*{-2cm}
\[
\hspace{-0.5cm}\epsfig{file=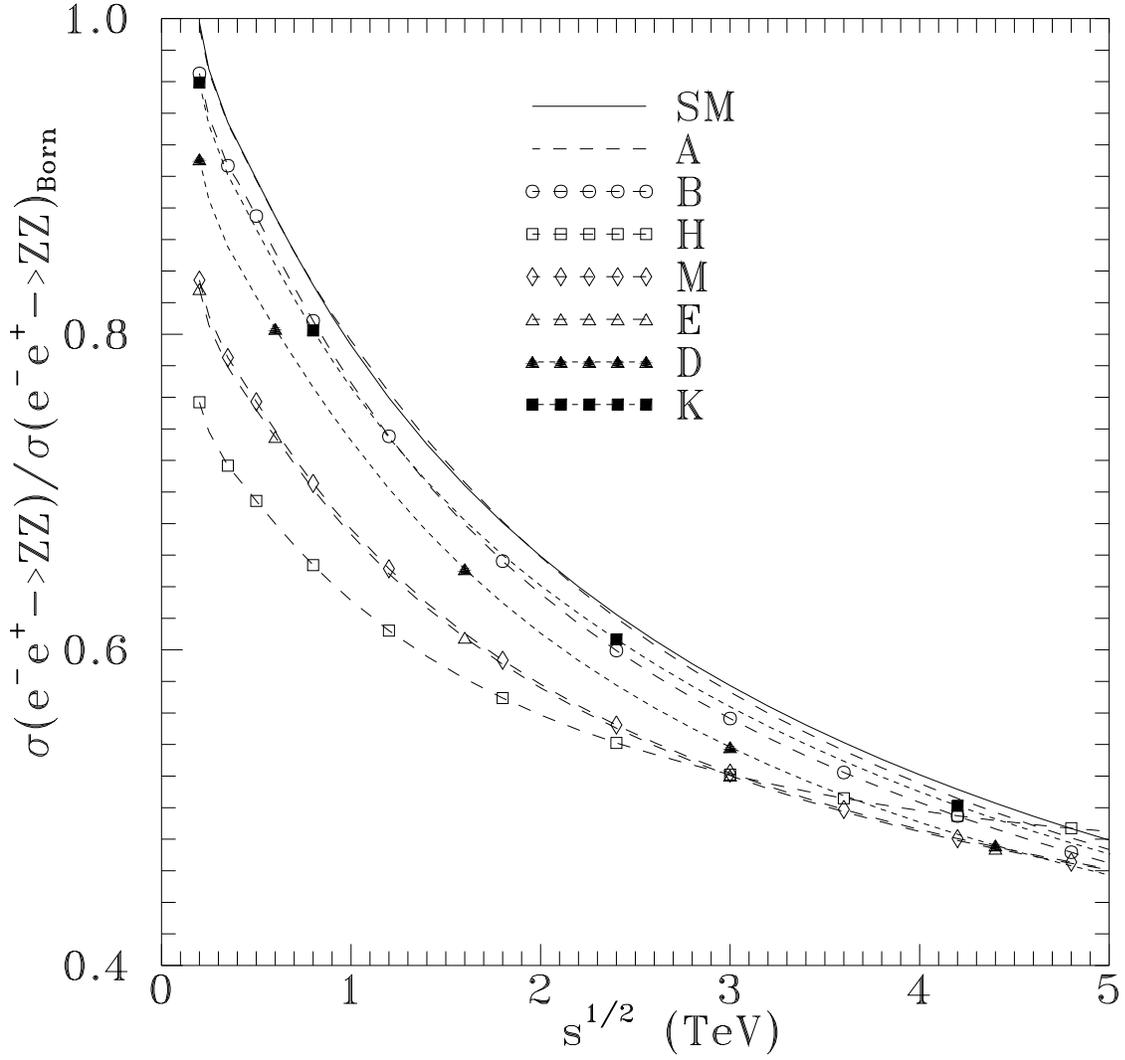,height=14cm}
\]
\vspace*{-0.5cm}
\caption[1]{The ratio of the integrated unpolarized
cross section $\sigma(e^-e^+\to Z Z )$ to the Born contribution,
for  SM, and a representative subset
of the benchmark MSSM models of \cite{Ellis-bench}.}
\label{ZZ-sig-fig}
\end{figure}

\clearpage
\newpage

\begin{figure}[p]
\vspace*{-2cm}
\[
\epsfig{file=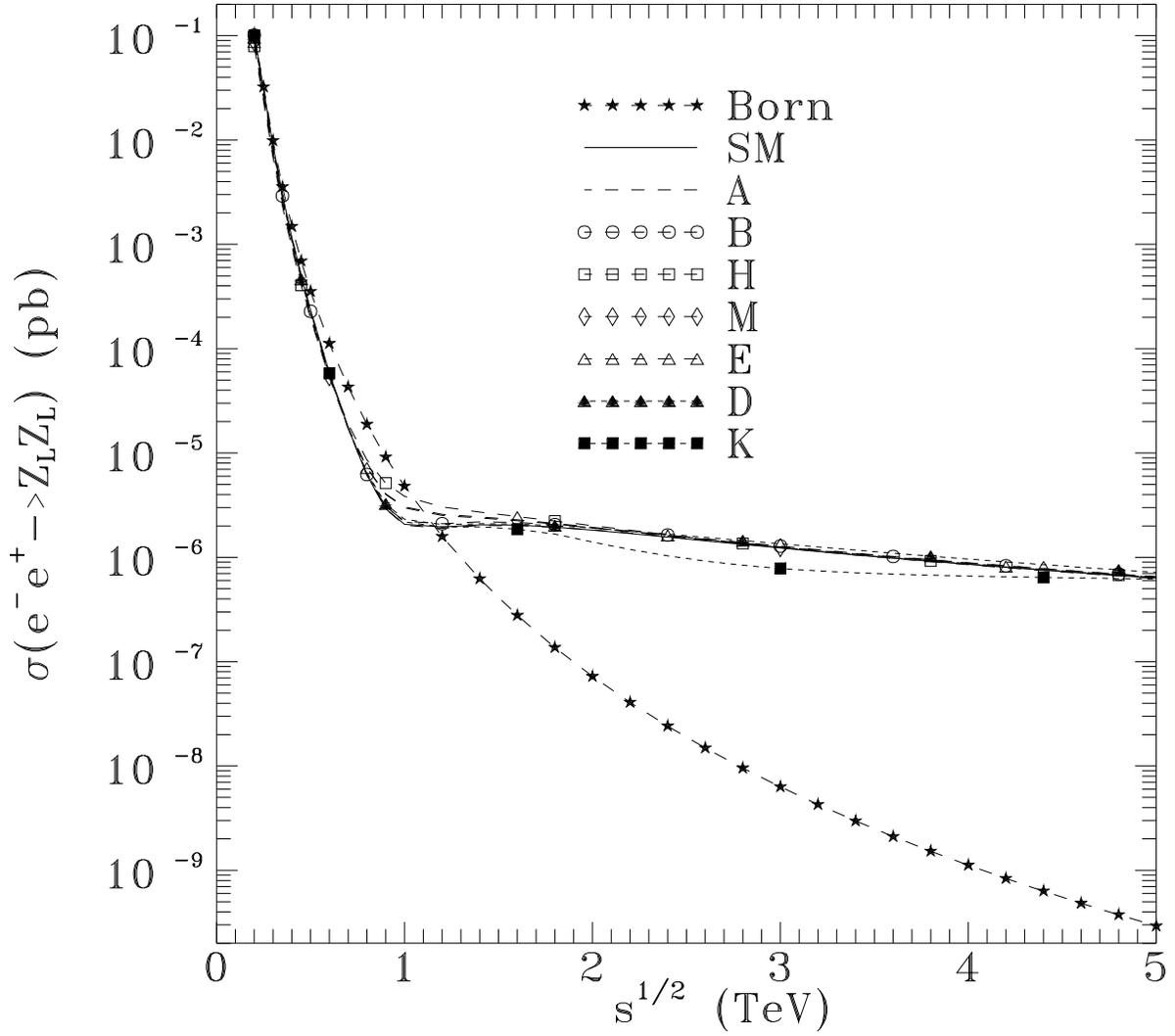,height=14cm}
\]
\vspace*{-0.5cm}
\caption[1]{The  integrated
cross section for $\sigma(e^-e^+\to Z _LZ_L )$,
as a function of the energy for
  SM and a representative subset
of the benchmark MSSM models of \cite{Ellis-bench}.
For comparison the Born contribution is also given.}
\label{ZZLL-sig-fig}
\end{figure}

\clearpage
\newpage

\begin{figure}[p]
\vspace*{-2cm}
\[
\hspace{0.1cm}\epsfig{file=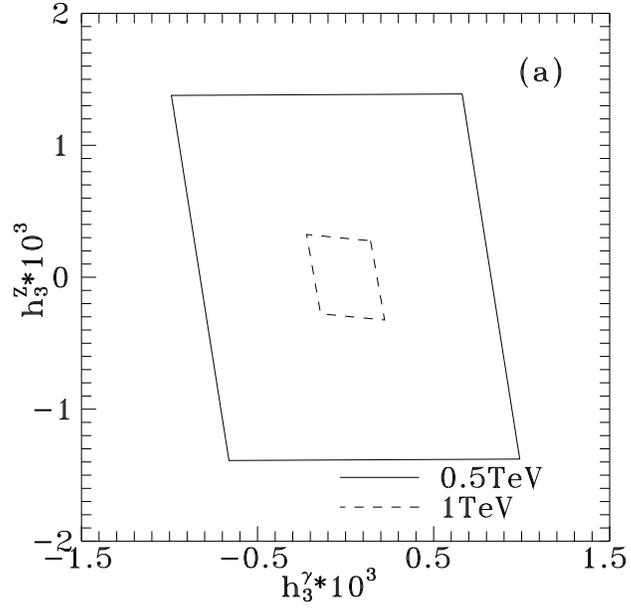,height=8cm}
\]
\vspace*{0.5cm}
\[
\hspace{0.1cm}\epsfig{file=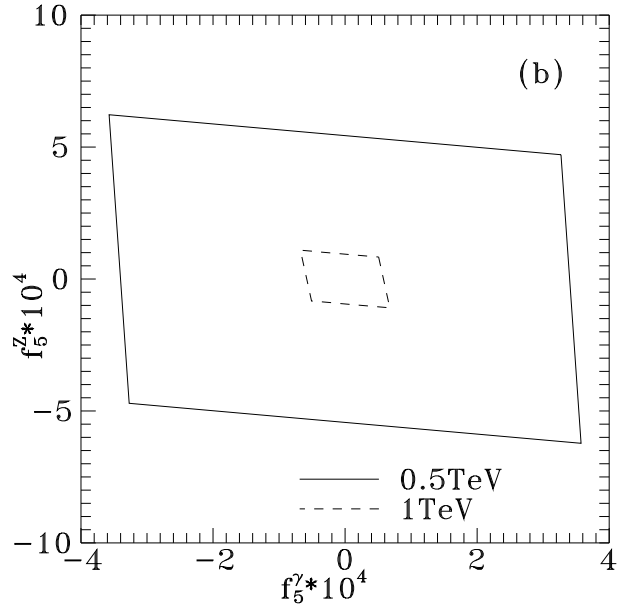,height=8cm}
\]
\vspace*{0.5cm}
\caption[1]{The NAGC limits  from $\sigma_{\rm unp}$ and $A_{LR}$
measurements  assuming an accuracy of $1\%$,
for the processes $e^+e^-\to Z\gamma$  (a), and
$e^+e^-\to ZZ$ (b).}
\label{NAGClim-fig}
\end{figure}

\end{document}